\documentclass[12pt,a4paper]{article}
\usepackage[utf8]{inputenc}
\usepackage{lmodern}
\usepackage[T1]{fontenc}
\usepackage[english]{babel}
\usepackage{ifpdf}
\usepackage[usenames,dvipsnames]{xcolor}
\usepackage{graphicx}
\usepackage[shortlabels]{enumitem}
\usepackage[numbers,sort]{natbib}
\usepackage[font=footnotesize,width=\textwidth]{caption}
\usepackage{booktabs}
\usepackage{hyperref}
\usepackage{uri}
\usepackage{tikz}
\usepackage{float}
\usepackage[top=1.8cm, bottom=2.8cm]{geometry}
\usepackage[parfill]{parskip}
\usepackage[affil-it]{authblk}
\usepackage[setbuildercolon]{matmatix}
\usepackage[normalem]{ulem}

\newcommand{\T}{\mathcal T}

\newcommand{\PP}{\mathbb P}
\newcommand{\blue}{\textcolor{black}}

\ifpdf
  \hypersetup{
    pdftitle={xxx}
  }
\fi

\hypersetup{colorlinks, allcolors = {MidnightBlue}}

\makeatletter
\renewcommand*{\defas}{%
\mathrel{\rlap{\raisebox{0.3ex}{$\m@th\cdot$}}\raisebox{-0.3ex}{$\m@th\cdot$}}=}
\makeatother

\makeatletter
\newtheorem*{rep@theorem}{\rep@title}
\newcommand{\newreptheorem}[2]{%
\newenvironment{rep#1}[1]{%
 \def\rep@title{#2 \ref*{##1}}%
 \begin{rep@theorem}}%
 {\end{rep@theorem}}}
\makeatother

\makeatletter
\DeclareRobustCommand\bigop[2][1]{%
  \mathop{\vphantom{\sum}\mathpalette\bigop@{{#1}{#2}}}\slimits@
}
\newcommand{\bigop@}[2]{\bigop@@#1#2}
\newcommand{\bigop@@}[3]{%
  \vcenter{%
    \sbox\z@{$#1\sum$}%
    \hbox{\resizebox{\ifx#1\displaystyle#2\fi\dimexpr\ht\z@+\dp\z@}{!}{$\m@th#3$}}%
  }%
}
\makeatother

\newtheorem{theorem}{Theorem}
\newtheorem{pro}[theorem]{Proposition}
\newreptheorem{pro}{Proposition}
\newtheorem{lem}[theorem]{Lemma}
\newreptheorem{lem}{Lemma}
\newtheorem{cor}[theorem]{Corollary}
\theoremstyle{definition}

\newcommand{\TT}{\mathbb{T}}

\title{\vspace{-1cm}
{Predicting the depth of the most recent common ancestor of a random sample of $k$ species:  the impact of phylogenetic tree shape}}

\author [1] {Michael Fuchs}
\author[2]{Mike Steel}

\affil[1]{Department of Mathematical Sciences, National Chengchi University, Taipei 116, Taiwan}
\affil[2]{Biomathematics Research Centre, University of Canterbury, Christchurch, New~Zealand}

\begin{document}

\maketitle

\begin{abstract}
We consider the following question: how close to the ancestral root of a phylogenetic tree is the most recent common ancestor of $k$ species randomly sampled from the tips of the tree? For trees having shapes predicted by the Yule--Harding model, it is known that the most recent common ancestor is likely to be close to (or equal to) the root of the full tree, even as $n$ becomes large (for $k$ fixed). However, this result does not extend to models  of tree shape that more closely describe phylogenies encountered in evolutionary biology.  We investigate the impact of tree shape (via the Aldous $\beta-$splitting model) to predict the number of edges that separate the most recent common ancestor of a random sample of $k$ tip species and the root of the parent tree they are sampled from.  Both exact and asymptotic results are presented. We also briefly consider a variation of the process in which a random number of tip species are sampled.
\end{abstract}


\section{Introduction}
A range of simple speciation--extinction processes predict the same distribution for the shape of phylogenetic trees (the classic and simple Yule--Harding (YH) model \cite{lam}). This model leads to a number of interesting predictions. One, in particular, was highlighted by Michael J. Sanderson \cite{san96} in 1996. He showed that sampling just 40 species at random from the tips of a large phylogenetic tree generated under the YH model is sufficient to ensure, with 95\% probability, that the most recent common ancestor (MRCA) of these 40 species at the tips the tree it was sampled from will exactly coincide with the global ancestral root of the tree. This is relevant for biologists wishing to estimate ancestral states near the origin of a large clade, since a small subset of the tip species of the clade may suffice for this task, particularly when a tree is not available for the entire clade. A remarkable feature of the prediction in \cite{san96} is that the number $n$ of leaves in the larger tree plays a vanishing role. More precisely,  \cite{san96} showed that the asymptotic probability (as $n$ grows) that the root of the parent tree and the MRCA of the $k$ sampled tip species coincide is  $1-\frac{2}{k+1}$. 

However, other models for describing phylogenetic tree shape can lead to quite different predictions. We focus here on the 1-parameter Aldous~$\beta$-splitting model, which includes the Yule--Harding model (when $\beta =0$), which tends to produce trees that are `overly balanced' compared with phylogenetic trees reconstructed from biological data \cite{blu06, moo97, ald01, hag15}. 

As observed by Aldous \cite{ald01} and others (e.g. \cite{hag15}) the $\beta$-splitting model with   $\beta = -1$  provides a reasonable description of the shape of many empirical trees in phylogenetic studies (particularly in comparison with  the Yule--Harding model). For the $\beta=-1$ model, we show that $k$ must grow as a power of $n$ (specifically $n^\alpha$) in order for the root of the parent tree and the root of the sampled subtree to coincide with probability $\alpha$.  For example, for a tree of 1000 tip species generated according to the $\beta=-1$ model, one needs to sample nearly 700 tip species in order for the MRCA to coincide with the root with probability 0.95 (in contrast to the 40 tips required for the Yule--Harding model). 

A more refined question than asking if the MRCA of the sample coincides with the root of the parent tree  $T$ is to ask `how close' the MRCA of the sampled tree is to the root of $T$, as measured by the number of edges between the two vertices. For the Yule--Harding model, this distance has earlier been shown to follow a geometric distribution as $n$ becomes large \cite{mck01}.
Here, we explore this distribution for other values of $\beta$, and show that it again is described by a geometric distribution for all $\beta >-1$, whereas the distribution for $\beta \leq -1$ is more complex. 

We also briefly consider a complementary sampling process. Rather than fixing the sample size $k$, we suppose that each species at the tips of $T$ is sampled independently with some fixed probability $p$. Thus, the number $K$ of sampled species across the $n$ tips of $T$ has the binomial distribution ${\rm Bin}(n,p)$. This model is relevant to biodiversity conservation under the simple `field of bullets' model of rapid extinction at the present \cite{rau93, nee97}. Under this  extinction model, one would like to estimate the extent to which the pruned tree captures the most ancestral parts of the original tree. In the Appendix, we derive analogous results (to those in the main part of the paper) when $k$ is replaced by $K$.


\blue{We end this section by noting that our approach and arguments are based on combinatorial enumeration and asymptotic methods. Several of the results presented here could likely also be derived using probabilistic techniques.}

\bigskip

\subsection{Definitions: Phylogenetic trees and models}\label{def-models}

For a set $X$ of species, a {\em phylogenetic $X$--tree $T$} is a tree for which $X$ is the set of vertices of in-degree 1 and out-degree 0 (the {\em leaves} of $T$) and for which every non-leaf vertex has out-degree 2.  Notice that a phylogenetic tree has a single vertex that has in-degree 0 and out-degree 2, referred to as the {\em root} of $T$. The set $\TT(X)$ of phylogenetic trees on a given set $X$ of size $n$ has size $(2n-3)!! = \prod_{j=1}^{n-1} (2j-1)$. If we ignore the labels of the leaves of a phylogenetic tree we obtain a {\em tree shape}.

\begin{figure}[ht]
    \centering
    \includegraphics[width=00.95\linewidth]{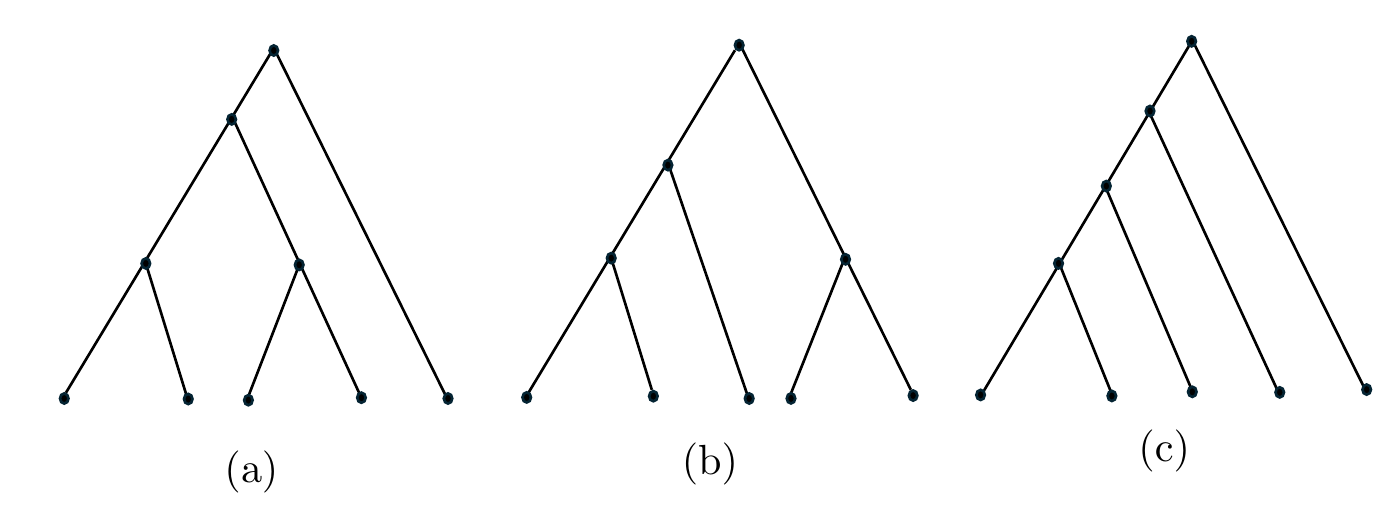}
    \vspace*{0.1cm}\caption{The three shapes of a phylogenetic tree on five leaves. For the Yule--Harding model ($\beta=0$), the tree shapes (a), (b), and (c) have probabilities 1/6, 1/2, and 1/3, respectively. For the PDA model ($\beta = -3/2)$, the corresponding tree shape probabilities are 1/7, 2/7 and 4/7; for the $\beta-$splitting model with $\beta = -1$, the tree shape probabilities are 9/55, 22/55 and 24/55.}
    \label{fig1}
\end{figure}

There are several models for randomly generating trees in $\TT(X)$. One such model is to simply select a tree uniformly at random from $\TT(X)$; this is referred to in evolutionary biology as the {\em PDA model} (here PDA refers to `proportional-to-distinguishable arrangements'). An alternative phylogenetic  model that is more closely based on an underlying evolutionary process is the {\em Yule--Harding} \cite{har} model.  In this model, one starts with a tree shape on two (unlabelled) leaves, and applies the following simple rule to build up a tree shape \blue{on the leaf label set $\{1, \ldots, n\}$}: Select uniformly at random a leaf of the tree shape so-far constructed and make it the parent of two new leaves. When there are $n$ leaves, these are then randomly assigned by the $n$ elemets of $X$. This discrete stochastic model leads to phylogenetic trees that tend to be more `balanced' than trees generated under the PDA model (see Fig.~\ref{fig1}). 

The Yule--Harding model  and the PDA model are special cases of a 1-parameter probability distribution on phylogenetic trees referred to as the {\em $\beta$-splitting} model of David Aldous \cite{ald96}. This  model recursively constructs a phylogenetic tree on $n$ leaves as follows. First, place $n$ points independently and uniformly at random on the interval $(0,1)$. Then split the $n$ points into two subsets of size $J \in \{1, \ldots, n-1\}$ and $n-J$ by cutting the interval $(0,1)$ according to a particular density $\pi_{n,j}$ (defined later and dependent on a parameter $\beta>-2$). This process is then repeated independently on each of these subsets (where $n$ is now replaced by $J$ and by $n-J$) and the process is continued until a tree on $n$ leaves results (these are then labelled randomly by elements of $X$). 

By varying the parameter $\beta$ (for $\beta >-2)$,  this model provides a way to generate trees of varying degrees of balance, including the PDA model (for $\beta = -3/2$) and the Yule--Harding model (for $\beta = 0$). The parameter $\beta=-1$ is also of particular interest, both mathematically (the model behaves differently than for other values of $\beta$) and for applications to evolutionary studies
(it provides an adequate description of the discrete shape of real phylogenetic trees, as first noted in \cite{ald01}). 
Fig.~\ref{fig1} describes how the three models of particular interest ($\beta = 0, -1, -3/2$) make different predictions concerning the shapes of trees with five leaves.

\vspace*{0.15cm}

\subsection{Depth of most recent common ancestors}

The {\em depth} of a vertex $v$ of $T$ is the number of edges in the directed path from the root of $T$ to $v$. Thus, $v$ has depth $0$ if and only if $v$ is the root of $T$.

\begin{figure}[ht]
    \centering
    \includegraphics[width=0.95\linewidth]{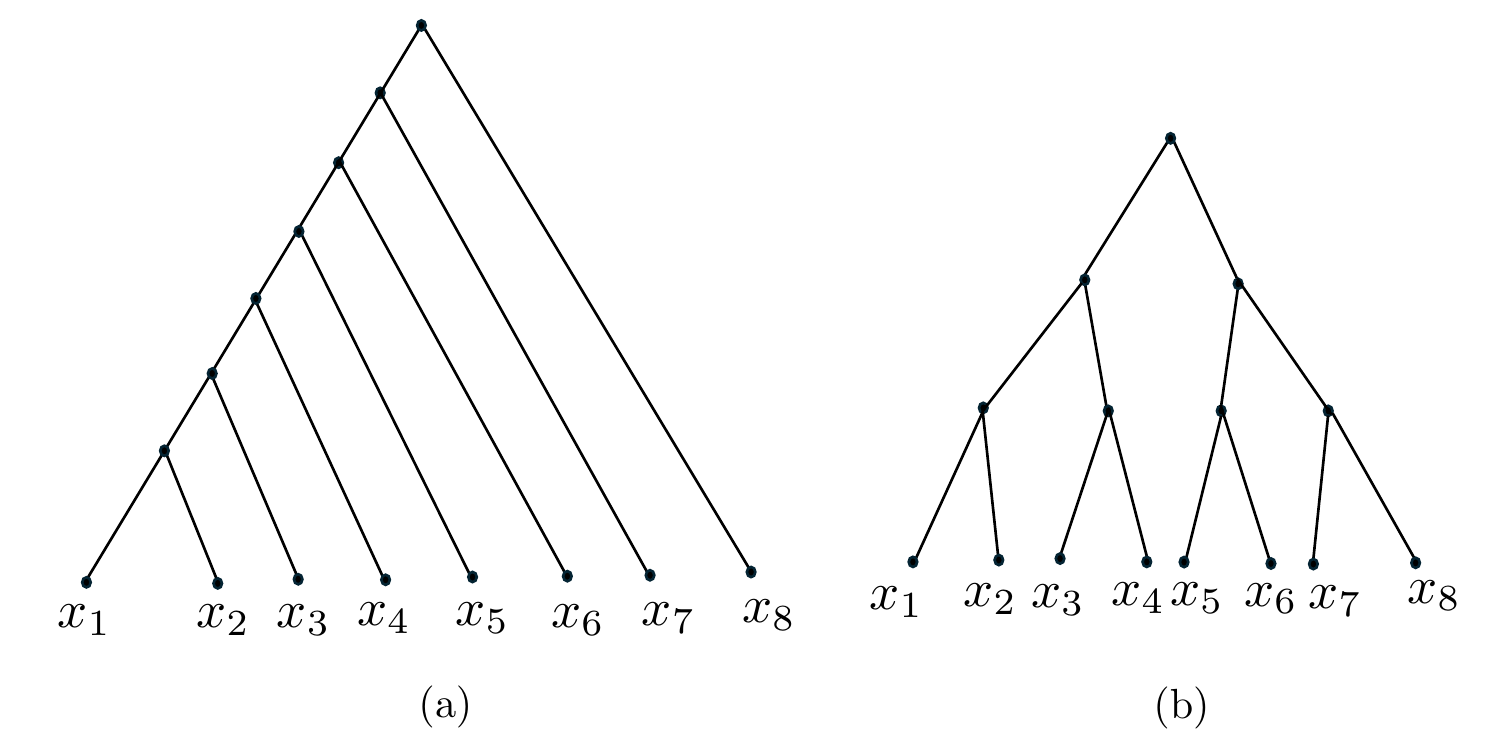}
    \vspace*{0.1cm}\caption{Two phylogenetic trees, one maximally  unbalanced (a), the other maximally balanced (b).  
    The MRCA of the set $S=\{x_2, x_5\}$ has depth 3 in tree (a) and depth 0 in tree (b). Also, if a set $S$ of three species is chosen uniformly at random from $\{x_1, \ldots, x_8\}$ then the probability that ${\rm MRCA}_T(S)$ coincides with the root of the tree is $1-\binom{7}{3}/\binom{8}{3}=\frac{3}{8}$ for tree (a) and $1-2\binom{4}{3}/\binom{8}{3}=\frac{6}{7}$ for tree (b).}
    \label{fig2}
\end{figure}

Given a phylogenetic $X$--tree $T$ and a subset $S$ of $X$, the {\em most recent common ancestor} of $S$ in $T$, denoted ${\rm MRCA}_T(S)$, is the (unique) vertex of $T$ that is (i) ancestral to each species in $S$, and (ii) has maximal depth for property (i).  This concept is illustrated in Fig.~\ref{fig2} for a fixed set $S$, and also when $S$ is a set of three leaves chosen uniformly at random from the eight leaves. 

\vspace*{0.25cm}

For a set $X$ of $n$ species, let  $D_{n,k}$ denote the depth of the  most recent common ancestor of a random subset of $X$ of size $k$ in a phylogenetic tree $\T$ on $X$ generated by the $\beta$-splitting model. 

Notice that there are two random processes at play here: the process that generates the tree, and the random sampling of $k$ leaves from this tree. Note also that $D_{n,k}=0$ if and only if the most recent common ancestor of the subset of $k$ leaves coincides with the root of $\T$.

\section{Main results}\label{results}
For each integer $m \geq 1$, let  $$H_m = \sum_{i=1}^m \frac{1}{i},$$
and let
\begin{equation}
\label{q-eq}
    q(\beta,k)=\frac{(\beta+2)\cdots(\beta+k)}{(2\beta+3)\cdots(2\beta+k+1)},
\end{equation}

where $k\geq 1$ is an integer and $\beta>-1$ is a real number.

Our first main result, Theorem~\ref{thm1}, presents exact and asymptotic expressions for the probability that the MRCA of $k$ randomly sampled leaves of a tree generated under the $\beta$-splitting model  coincides with the root of the tree.  The cases $\beta = -1$ and $\beta \neq -1$ need to be treated separately. 

\bigskip 
\begin{theorem}
\label{thm1}
\mbox{}
\begin{itemize}
    \item[(i)] For $\beta=-1$, and all $n\geq 1$ and $2\leq k\leq n$,
    \begin{equation}\label{beta=-1}
{\mathbb P}(D_{n,k}=0)=\frac{H_{k-1}}{H_{n-1}},
\end{equation}
which tends to $0$ (at logarithmic speed) for fixed $k$. However, if $k$ is allowed to grow  as a sublinear power of $n$ (i.e.  $k\sim n^{\alpha}$ with $0<\alpha\leq 1$), then:
\[
{\mathbb P}(D_{n,k}=0)\sim\frac{\log(n^{\alpha})}{\log n}=\alpha.
\]
    \item[(ii)] For $\beta \neq -1$,
\begin{equation}
\label{betanotequal}
{\mathbb P}(D_{n,k}=0)=1-\frac{2(\beta+1)\cdots(\beta+k)}{k!\binom{n}{k}}\cdot\frac{\binom{n+2\beta+1}{n-k}-\binom{n+\beta}{n-k}}
{\binom{n+2\beta+1}{n}-2\binom{n+\beta}{n}}.
\end{equation}
Moreover, for $k$ fixed, 
\[
\lim_{n\rightarrow\infty}{\mathbb P}(D_{n,k}=0)=\begin{cases} 1-q(\beta,k)>0,&\text{if}\ \beta>-1;\\ 0,&\text{if}\ -2<\beta<-1\end{cases}
\]
and for the second case ($-2<\beta<-1$), if $k$ grows with $n$, we have:
\[
\lim_{n\rightarrow\infty}{\mathbb P}(D_{n,k}=0)=\begin{cases} 0,&\text{if}\ k=o(n);\\ c^{-\beta-1},&\text{if}\ k\sim cn.\end{cases}.
\]
\end{itemize}
\end{theorem}

{\bf Remark:}
Equations (\ref{beta=-1}) and (\ref{betanotequal}) also hold for $k=1$ as, in this case, the expressions for ${\mathbb P}(D_{n,k}=0)$ both equal $0$.

\bigskip

To illustrate Theorem~\ref{thm1}, the following table lists  the smallest value of $k$ for which ${\mathbb P}(D_{n,k}=0)\geq 0.95$ on trees of an increasing number ($n$) of leaves, generated under the Yule--Harding model $(\beta = 0)$, the $\beta=-1$  model, and the PDA model (for which $\beta = -3/2).$ Note that for the Yule--Harding model, 
\[
{\mathbb P}(D_{n,k}=0)= 1-2\frac{n-k}{(n-1)(k+1)}
\]
by Part (ii).

\vspace*{0.2cm}

\begin{center}
\begin{tabular}{ c || c | c | c | c | c |c }
 $n$ & $10$ & $10^2$ & $10^3$ & $10^4$ & $10^5$ & $10^6$\\ 
 \hline\hline
 $\beta=0$  & 8 & 29 & 38 & 39 & 39 & 39\\
 \hline
 $\beta=-1$  & 9 & 78 & 688 & 6131 & 54635 & 486930\\
 \hline
 $\beta=-3/2$ & 10 & 91 & 903 & 9026 & 90251 & 902501 \\
\end{tabular}
\end{center}

\bigskip

Our second main result, Theorem~\ref{thm2}, describes the asymptotic distribution of $D_{n,k}$ as $n$ grows (with $k$ fixed) for $\beta \geq -1$, and a special case with $-2<\beta <-1$ (namely, $\beta=-3/2$, which corresponds to the PDA model; for the remaining cases, we have a conjecture which is presented in the conclusion). 

In the following theorem, $\stackrel{d}{\longrightarrow}$ refers to convergence in distribution as $n \rightarrow \infty$.

\bigskip
\begin{theorem}
\label{thm2}
\mbox{}
\begin{itemize}
 \item[(i)]
 For $\beta >-1$, and $k\geq 2$ fixed,  as
 $n\rightarrow\infty$, 
\[
D_{n,k} \stackrel{d}{\longrightarrow} G_k,
\]
where $G_k$ has a geometric distribution, with 
${\mathbb P}(G_k = r) = (1-q(\beta, k))q(\beta, k)^{r}$ for 
$r \geq 0$.
    \item[(ii)] 
    For $\beta = -1$ and $k\geq 2$ fixed,
   as $n\rightarrow\infty$,
\[
\frac{H_{k-1}D_{n,k}}{\log n}\stackrel{d}{\longrightarrow}{\rm Exp}(1),
\]
where ${\rm Exp}(1)$ is the standard exponential distribution.
\item[(iii)] 
For $\beta = -3/2$ and $k\geq 1$ fixed, as $n\rightarrow\infty$,
\[
\frac{D_{n,k}}{\sqrt{n}}\stackrel{d}{\longrightarrow} D_k,
\]
where $D_k$ is a distribution which is uniquely characterized by its moments sequence
\begin{equation}\label{mom-Dk}
{\mathbb E}(D_k^m)=\frac{m!C_{k-1}4^{1-k}k!\sqrt{\pi}}{\Gamma(k+(m-1)/2)},
\end{equation}
where $C_{k-1}$ denotes the $(k-1)$-st Catalan number, and $\Gamma$ is the Gamma function.
\end{itemize}
\end{theorem}

\bigskip

Note that the restriction that $k\geq 2$ in Parts (i) and (ii) is necessary, since if $k=1$, the geometric and exponential distributions (respectively) are replaced by a normal distribution: For $\beta=-1$, see Theorem~1.7 in \cite{ald24}; for $\beta>-1$, a normal distribution is suggested by the results in \cite{ald96} and the (known) limiting distribution result for $\beta=0$ from, e.g., Section~2.4 in \cite{Ma}.

\section{Proof of Theorem~\ref{thm1}}

In this section, we prove Theorem~\ref{thm1}. As this theorem contains both exact and asymptotics results, we separate our considerations into two subsections.

\subsection{Exact evaluation}

First, recall from \cite{ald96} that the density $\pi_{n,j}$ in the $\beta$-splitting model (see Section~\ref{def-models}) is given by
\begin{equation}\label{split-probab}
\pi_{n,j}=\frac{1}{c_{n}(\beta)}\frac{\Gamma(j+\beta+1)\Gamma(n-j+\beta+1)}{j!(n-j)!},\qquad (1\leq j\leq n-1),
\end{equation}
where the normalization constant $c_n(\beta)$ is such that $\sum_{j}\pi_{n,j}=1$. Consequently,
\[
c_n(\beta)=\sum_{j=1}^{n-1}\frac{\Gamma(j+\beta+1)\Gamma(n-j+\beta+1)}{j!(n-j)!}.
\]
The starting point of our analysis is the following (simple) lemma.

\bigskip

\begin{lem}
We have,
\begin{equation}\label{Dnk-gen}
{\mathbb P}(D_{n,k}\geq 1)=\sum_{j=1}^{n-1}\frac{\binom{j}{k}+\binom{n-j}{k}}{\binom{n}{k}}\pi_{n,j}=\frac{2}{\binom{n}{k}}\sum_{j=1}^{n-1}\binom{j}{k}\pi_{n,j}.
\end{equation}
\end{lem}
\begin{proof}
The depth of the most recent common ancestor is larger than $0$ if and only if all $k$ sampled leaves belong to either $J$ or $n-J$ in the definition of the $\beta$-splitting model. Thus, the probability that this happens if $\vert J\vert=j$ equals
\[
\frac{\binom{j}{k}+\binom{n-j}{k}}{\binom{n}{k}}.
\]
Multiplying this ratio by the probability that $\vert J\vert=j$ (which is given by $\pi_{n,j}$) and \blue{summing}  over $j$ gives the desired result.
\end{proof}

We now consider the cases $\beta=-1$ and $\beta\ne -1$ separately. 

\paragraph{\blue{Case I: $\beta=-1$.}} First, note that
\[
\sum_{j=1}^{n-1}\frac{\Gamma(j)\Gamma(n-j)}{j!(n-j!)}=\frac{1}{n}\sum_{j=1}^{n-1}\left(\frac{1}{j}+\frac{1}{n-j}\right)=
\frac{2H_{n-1}}{n}
\]
which implies 
\[
c_n(\beta)=\frac{2H_{n-1}}{n}.
\]
Plugging this into (\ref{split-probab}) yields
\[
\pi_{n,j}=\frac{n}{2H_{n-1}}\cdot\frac{1}{j(n-j)},\qquad (1\leq j\leq n-1).
\]
Thus, (\ref{Dnk-gen}) becomes
\begin{align}
{\mathbb P}(D_{n,k}\geq 1)&=\frac{n}{\binom{n}{k}H_{n-1}}\sum_{j=1}^{n-1}\binom{j}{k}\frac{1}{j(n-j)}\nonumber\\
&=\frac{n}{k\binom{n}{k}H_{n-1}}\sum_{j=1}^{n-1}\binom{j-1}{k-1}
\frac{1}{n-j}.\label{Dnk-crit}
\end{align}
The latter sum can be computed as follows. 

\bigskip

\begin{lem}\label{eval}
We have,
\[
\sum_{j=1}^{n-1}\binom{j-1}{k-1}\frac{1}{n-j}=(H_{n-1}-H_{k-1})\binom{n-1}{k-1}.
\]
\end{lem}
\begin{proof}
The sum is a convolution and thus by using ordinary generating functions:
\[
\sum_{j=1}^{n-1}\binom{j-1}{k-1}\frac{1}{n-j}=[z^n]\left(\sum_{j\geq 1}\binom{j-1}{k-1}z^j\right)\left(\sum_{j\geq 1}\frac{z^j}{j}\right).
\]
Note that
\[
\sum_{j\geq 1}\frac{z^j}{j}=\log\left(\frac{1}{1-z}\right)
\]
and
\begin{align*}
\sum_{j\geq 1}\binom{j-1}{k-1}z^j&=\frac{z}{(k-1)!}\sum_{j\geq 0}j(j-1)\cdots(j-k+2)z^j\\
&=\frac{z^k}{(k-1)!}\left(\frac{1}{1-z}\right)^{(k-1)}=\frac{z^k}{(1-z)^k}.
\end{align*}
Thus,
\begin{align*}
\sum_{j=1}^{n-1}\binom{j-1}{k-1}\frac{1}{n-j}&=[z^n]\frac{z^k}{(1-z)^k}\log\left(\frac{1}{1-z}\right)\\
&=[z^{n-k}]\frac{1}{(1-z)^k}\log\left(\frac{1}{1-z}\right).
\end{align*}
The result follows from this by using the expansion:
\[
\frac{1}{(1-z)^{k+1}}\log\left(\frac{1}{1-z}\right)=\sum_{\ell\geq 0}(H_{\ell+k-1}-H_{k-1})\binom{\ell+k-1}{\ell}z^{\ell}.
\]
This concludes the proof.
\end{proof}

Substituting the identity from Lemma~\ref{eval} into (\ref{Dnk-crit}), we have the following equation which establishes the exact result from Theorem~\ref{thm1}, Part (i).

\bigskip

\begin{cor}\label{cor-crit}
\blue{For $\beta = -1$, we have:}
\[
{\mathbb P}(D_{n,k}\geq 1)=1-\frac{H_{k-1}}{H_{n-1}}.
\]
\end{cor}
\begin{proof}
We have,
\[
{\mathbb P}(D_{n,k}\geq 1)=\frac{n\binom{n-1}{k-1}}{k\binom{n}{k}H_{n-1}}\left(H_{n-1}-H_{k-1}\right)=1-\frac{H_{k-1}}{H_{n-1}},
\]
where we used that
\[
\frac{n}{k}\binom{n-1}{k-1}=\binom{n}{k}.
\]
This establishes the claim.
\end{proof}
\paragraph{\blue{Case II: $\beta\ne -1$.}} In this case, we rewrite (\ref{Dnk-gen}) as
\[
{\mathbb P}(D_{n,k}\geq 1)=\frac{2\Gamma(\beta+1)^2}{k!\binom{n}{k}c_n(\beta)}\sum_{j=1}^{n-1}j(j-1)\cdots(j-k+1)\binom{j+\beta}{j}\binom{n-j+\beta}{n-j}.
\]

\newpage
In order to simplify the sum, recall that
\[
\sum_{j\geq 1}\binom{j+\beta}{j}z^j=(1-z)^{-\beta-1}-1.
\]
Thus,
\begin{equation}\label{Dnk-non-crit}
{\mathbb P}(D_{n,k}\geq 1)=\frac{2\Gamma(\beta+1)^2}{k!\binom{n}{k}c_n(\beta)}[z^n]z^{k}\left((1-z)^{-\beta-1}-1\right)^{(k)}\left((1-z)^{-\beta-1}-1\right).
\end{equation}
Note that
\[
\left((1-z)^{-\beta-1}-1\right)^{(k)}=(\beta+1)\cdots(\beta+k)(1-z)^{-\beta-1-k}
\]
and consequently:
\begin{align*}
[z^n]z^{k}&\left((1-z)^{-\beta-1}-1\right)^{(k)}\left((1-z)^{-\beta-1}-1\right)\\
&=(\beta+1)\cdots(\beta+k)[z^{n-k}](1-z)^{-2\beta-2-k}-(1-z)^{-\beta-1-k}\\
&=(\beta+1)\cdots(\beta+k)\left(\binom{n+2\beta+1}{n-k}-\binom{n+\beta}{n-k}\right).
\end{align*}
Likewise,
\begin{align}
\frac{c_n(\beta)}{\Gamma(\beta+1)^2}&=\sum_{j=1}^{n-1}\binom{j+\beta}{j}\binom{n-j+\beta}{j}\nonumber\\
&=[z^n]\left((1-z)^{-\beta-1}-1\right)^2=\binom{n+2\beta+1}{n}-2\binom{n+\beta}{n}.\label{norm-constant}
\end{align}
Plugging everything into (\ref{Dnk-non-crit}) gives the following, which implies the exact result from Theorem~\ref{thm1}, Part (ii).

\bigskip 

\begin{cor}\label{cor-non-crit}
\blue{For $\beta\ne -1$, we have:}
\[
{\mathbb P}(D_{n,k}\geq 1)=\frac{2(\beta+1)\cdots(\beta+k)}{k!\binom{n}{k}}\cdot\frac{\binom{n+2\beta+1}{n-k}-\binom{n+\beta}{n-k}}
{\binom{n+2\beta+1}{n}-2\binom{n+\beta}{n}}.
\]
\end{cor}

\subsection{Asymptotic evaluation}

Here, we are interested in the limit as $n$ tends to infinity of the probability that $D_{n,k}=0$. \blue{Again, we split our considerations into two cases.}

\paragraph{\blue{Case I: $\beta=-1$.}} First, from Corollary~\ref{cor-crit}, we have
\[
{\mathbb P}(D_{n,k}=0)=\frac{H_{k-1}}{H_{n-1}}.
\]
This tends to $0$ at logarithmic speed for fixed $k$. On the other hand, if $k$ depends on $n$ such that $k\sim n^{\alpha}$ with $0<\alpha\leq 1$, then
\[
{\mathbb P}(D_{n,k}=0)\sim\frac{\log(n^{\alpha})}{\log n}=\alpha,
\]
as stated in the asymptotic part of Theorem~\ref{thm1}, Part (i).


\paragraph{\blue{Case II: $\beta\ne -1$.}}Next, we consider $\beta\ne-1$. Here, we rewrite the result from Corollary~\ref{cor-non-crit} as follows:
\begin{align}
{\mathbb P}(D_{n,k}\geq 1)&=2(\beta+1)\!\cdots\!(\beta+k)\frac{\displaystyle\frac{\Gamma(n+2\beta+2)}{\Gamma(k+2\beta+2)\Gamma(n+1)}-\frac{\Gamma(n+\beta+1)}
{\Gamma(k+\beta+1)\Gamma(n+1)}}{\displaystyle\frac{\Gamma(n+2\beta+2)}{\Gamma(2\beta+2)\Gamma(n+1)}-2\frac{\Gamma(n+\beta+1)}{\Gamma(\beta+1)\Gamma(n+1)}}
\nonumber\\[5pt]
&=\frac{{\displaystyle\frac{q(\beta,k)}{\Gamma(2\beta+2)}\frac{\Gamma(n+2\beta+2)}{\Gamma(n+1)}-2\frac{\Gamma(n+\beta+1)}
{\Gamma(\beta+1)\Gamma(n+1)}}}{\displaystyle\frac{\Gamma(n+2\beta+2)}{\Gamma(2\beta+2)\Gamma(n+1)}-2\frac{\Gamma(n+\beta+1)}{\Gamma(\beta+1)\Gamma(n+1)}},
\label{probab-non-crit}
\end{align}
where $q(\beta,k)$ is as in \blue{Eqn.~\ref{q-eq} of} Section~\ref{results}. Consider first the denominator whose two terms have the following first-order expansions:
\begin{equation}\label{diff-asymp}
\frac{\Gamma(n+2\beta+2)}{\Gamma(2\beta+2)\Gamma(n+1)}\sim\frac{n^{2\beta+1}}{\Gamma(2\beta+2)}\qquad\text{and}\qquad
\frac{\Gamma(n+\beta+1)}{\Gamma(\beta+1)\Gamma(n+1)}\sim\frac{n^{\beta}}{\Gamma(\beta+1)}.
\end{equation}
Note that the first term is asymptotically dominant if $\beta>-1$ and the second term is asymptotically dominant if $-2<\beta<-1$. Thus,  for $\beta>-1$, we have
\begin{equation}
\label{eqx}
    \lim_{n\rightarrow\infty}{\mathbb P}(D_{n,k}\geq 1)=q(\beta,k)<1
\end{equation}
and for $-2<\beta<-1$
\[
\lim_{n\rightarrow\infty}{\mathbb P}(D_{n,k}\geq 1)=1.
\]
By combining both results, we have:
\begin{equation}\label{two-cases}
\lim_{n\rightarrow\infty}{\mathbb P}(D_{n,k}=0)=\begin{cases} 1-q(\beta,k)>0,&\text{if}\ \beta>-1;\\ 0,&\text{if}\ -2<\beta<-1;\end{cases}
\end{equation}
as stated in the asymptotic part of Theorem~\ref{thm1}, Part (ii).

The second \blue{part of (\ref{two-cases})} is akin to the case $\beta=-1$ and thus again raises the following question: if $k$ grows to infinity with $n$, how fast does $k$ have to grow so that the probability becomes strictly positive? In order to answer this, we have to refine the above computations. First, note that for the denominator of (\ref{probab-non-crit}), we have
\[
\frac{\Gamma(n+2\beta+2)}{\Gamma(2\beta+2)\Gamma(n+1)}-2\frac{\Gamma(n+\beta+1)}{\Gamma(\beta+1)\Gamma(n+1)}
=-\frac{2}{\Gamma(\beta+1)}n^{\beta}\left(1+{\mathcal O}(n^{\beta+1})\right).
\]
Moreover, for the two terms in the numerator of (\ref{probab-non-crit}), we have
\[
-2\frac{\Gamma(n+\beta+1)}{\Gamma(\beta+1)\Gamma(n+1)}=-\frac{2}{\Gamma(\beta+1)}n^{\beta}\left(1+{\mathcal O}(n^{-1})\right)
\]
and
\begin{align*}
\frac{q(\beta,k)}{\Gamma(2\beta+2)}\frac{\Gamma(n+2\beta+2)}{\Gamma(n+1)}&=\frac{2}{\Gamma(\beta+1)}\frac{\Gamma(k+\beta+1)}{\Gamma(k+2\beta+2)}
\frac{\Gamma(n+2\beta+2)}{\Gamma(n+1)}\\
&=\frac{2}{\Gamma(\beta+1)}n^{\beta}\left(\frac{k}{n}\right)^{-\beta-1}\left(1+{\mathcal O}(k^{-1})\right).
\end{align*}
Overall,
\begin{align}
{\mathbb P}(D_{n,k}\geq 1)&=1+{\mathcal O}(n^{\beta+1})-\left(\frac{k}{n}\right)^{-\beta-1}(1+{\mathcal O}(k^{-1})+{\mathcal O}(n^{\beta+1}))\nonumber\\
&=1-\left(\frac{k}{n}\right)^{-\beta-1}+{\mathcal O}(n^{\beta+1}).\label{beta<-1}
\end{align}
From this, we obtain the following result:
\[
\lim_{n\rightarrow\infty}{\mathbb P}(D_{n,k}=0)=\begin{cases} 0,&\text{if}\ k=o(n);\\ c^{-\beta-1},&\text{if}\ k\sim cn.\end{cases}.
\]
This is the last claim from the asymptotic part of Theorem~\ref{thm1}, Part (ii).

\section{Proof of Theorem~\ref{thm2}}\label{proof-thm2}

In this section, we prove Theorem~\ref{thm2}. We start with the cases $\beta=-3/2$ \blue{(the PDA model)} and $\beta=-1$ in Section~\ref{ll-two-cases}; the proof for $\beta>-1$ is presented in Section~\ref{ll-geo} and is quite different. For the former two cases, the starting point is the same, namely, we deduce the result from the following distributional recurrence for $D_{n,k}$ (which, in fact, holds for {\it all} cases of $\beta$):
\begin{equation}\label{dis-rec}
(D_{n,k}\vert I_n=j)\stackrel{d}{=}\begin{cases} \blue{D_{j,k}}+1,&\text{with probability}\ \binom{j}{k}/\binom{n}{k};\\
\blue{D_{n-j,k}}+1,&\text{with probability}\ \binom{n-j}{k}/\binom{n}{k};\\ 0,&\text{otherwise.}\end{cases}
\end{equation}
Here, $I_n$ denotes the (random) size of $J$ whose distribution is $\pi_{n,j}$, i.e., ${\mathbb P}(I_n=~j)=\pi_{n,j}$; see the definition of the $\beta$-splitting model in Section~\ref{def-models}.

This recurrence is explained as follows: if $\vert J\vert=j$, then the depth of the recent common ancestor to the root is increased by $1$ if and only if all the sampled leaves are chosen either from $J$ or from $n-J$ (which happens with probability $\binom{j}{n}/\binom{n}{k}$ and $\binom{n-j}{k}/\binom{n}{k}$, respectively); otherwise, the depth equals to $0$.

\subsection{Asymptotic distribution for $\beta=-3/2$ and $\beta=-1$}\label{ll-two-cases}

\paragraph{\blue{The PDA model}.} For $\beta=-\frac{3}{2}$, we use generating function techniques in combination with the method of singularity analysis (see Chapter VI in \cite{FlSe}), \blue{and the following result (Theorem 30.2 of \cite{bil79}):}

\vspace*{-0.5cm}
\blue{\begin{lem}
    \label{bil}
        Suppose that the distribution of a random variable is determined by its moments, and that random variables $X_n$ have moments of all orders, and that $\lim_{n\rightarrow \infty}{\mathbb E}(X_n^r) = {\mathbb E}(X^r])$ for $r=1, 2, \ldots$. Then $X_n$ converges in distribution to $X$.
\end{lem}}

First observe that (\ref{split-probab}) can be rewritten as
\[
\pi_{n,j}=\frac{C_{j-1}C_{n-1-j}}{C_{n-1}},\qquad (1\leq j\leq n-1),
\]
where $C_{n-1}$ denotes the $n-1$-st Catalan number with generating function:
\[
\sum_{n\geq 1}C_{n-1}z^n=\frac{1-\sqrt{1-4z}}{2}.
\]
Thus, from (\ref{dis-rec}), we have for the $m$-th moment
\[
{\mathbb E}(D_{n,k}^{m})=\frac{2}{\binom{n}{k}}\sum_{j=1}^{n-1}\binom{j}{k}{\mathbb E}(D_{j,k}+1)^m\frac{C_{j-1}C_{n-1-j}}{C_{n-1}}
\]
or equivalently, by multiplying by $C_{n-1}\binom{n}{k}$ and using the binomial theorem,
\begin{align*}
C_{n-1}\binom{n}{k}{\mathbb E}(D_{n,k}^{m})=2\sum_{j=1}^{n-1}&C_{j-1}\binom{j}{k}{\mathbb E}(D_{j,k}^m)C_{n-1-j}\\
&+\sum_{\ell=1}^{m}2\binom{m}{\ell}\sum_{j=1}^{n-1}C_{j-1}\binom{j}{k}{\mathbb E}(D_{j,k}^{m-\ell})C_{n-1-j}.
\end{align*}
Set
\[
D^{[m]}(z):=\sum_{n\geq 1}C_{n-1}\binom{n}{k}{\mathbb E}(D_{n,k}^{m})z^n.
\]
The above recurrence can then be translated into the following algebraic equation
\[
D^{[m]}(z)=(1-\sqrt{1-4z})D^{[m]}(z)+\sum_{\ell=1}^{m}\binom{m}{\ell}D^{[m-\ell]}(z)(1-\sqrt{1-4z})
\]
which has the solution
\begin{equation}\label{exact-sol}
D^{[m]}(z)=\sum_{\ell=1}^{m}\binom{m}{\ell}D^{[m-\ell]}(z)((1-4z)^{-1/2}-1).
\end{equation}
From this by induction, we obtain the following result, where a function $f(z)$ is called $\Delta$-analytic at $z_0$ if $f(z)$ is analytic in a domain
\[
\Delta=\{z\in{\mathbb C}\ :\ \vert z\vert<\vert z_0\vert+\epsilon,\ \vert\arg(z-z_0)\vert>\phi_0\}
\]
for some $\epsilon>0$ and $0<\phi_0<\pi/2$; see  Definition VI.1 in \cite{FlSe}.

\bigskip

\begin{pro}
For $m\geq 1$, $D^{[m]}(z)$ is $\Delta$-analytic at $1/4$ with expansion:
\[
D^{[m]}(z)\sim\frac{m!C_{k-1}4^{-k}}{(1-4z)^{k+(m-1)/2}},\qquad (z\rightarrow 1/4).
\]
\end{pro}

\begin{proof}
First, for $m=1$, we have
\[
D^{[1]}(z)=D^{[0]}(z)((1-4z)^{-1/2}-1),
\]
where
\[
D^{[0]}(z)=\sum_{n\geq 1}C_{n-1}\binom{n}{k}z^{n}=\frac{z^k}{k!}\left(\frac{1-\sqrt{1-4z}}{2}\right)^{(k)}=C_{k-1}z^k(1-4z)^{1/2-k}.
\]
Thus, as $z\rightarrow 1/4$,
\[
D^{[1]}(z)\sim\frac{C_{k-1}4^{-k}}{(1-4z)^{k}}
\]
which proves the claim for $m=1$.

Next, by induction and (\ref{exact-sol}), as $z\rightarrow 1/4$,
\[
D^{[m]}(z)\sim mD^{[m-1]}(z)(1-4z)^{-1/2}
\]
from which the claim follows.
\end{proof}

By the transfer theorems of singularity analysis (see Section VI.3 in \cite{FlSe}), we obtain the following limit law result; see Theorem~\ref{thm2}, Part (iii).

\bigskip

\begin{theorem}\label{limit-law-subcrit}
As $n\rightarrow\infty$,
\[
\frac{D_{n,k}}{\sqrt{n}}\stackrel{d}{\longrightarrow} D_k,
\]
where $D_k$ is a distribution which is uniquely characterized by its moments sequence
\begin{equation}\label{mom-Dk}
{\mathbb E}(D_k^m)=\frac{m!C_{k-1}4^{1-k}k!\sqrt{\pi}}{\Gamma(k+(m-1)/2)}.
\end{equation}
\end{theorem}
\begin{proof}
By the transfer theorems,
\[
[z^n]D^{[m]}(z)\sim\frac{m!C_{k-1}4^{n-k}}{\Gamma(k+(m-1)/2)}n^{k+(m-3)/2}.
\]
From this, by the asymptotic expansions,
\[
C_{n-1}\sim\frac{4^n}{\sqrt{\pi n^3}}\qquad\text{and}\qquad \binom{n}{k}\sim\frac{n^k}{k!},
\]
we obtain that
\[
{\mathbb E}(D_{n,k}^m)\sim\frac{m!C_{k-1}4^{1-k}k!\sqrt{\pi}}{\Gamma(k+(m-1)/2)}n^{m/2}.
\]
As the multiplicative factor on the right-hand side is the moment sequence of a unique distribution, the result follows \blue{from Lemma~\ref{bil}.}
\end{proof}
{\bf Remark:} For $k=1$, Equation (\ref{mom-Dk}) describes the moments of $2\sqrt{2}e(r)$, where $\{e(t), 0\leq t\leq 1\}$ is the \blue{normalized} Brownian excursion and $r$ is a point from $[0,1]$ picked uniformly at random. More generally, the distribution of $D_k$ is a three-parameter Mittag-Leffler distribution ${\rm ML}(\alpha,\beta,\gamma)$ with $\alpha=\beta=1/2$ and $\gamma=k-1$; see Definition~3.11 and Lemma~3.12 in \cite{BaKuWa}.

\paragraph{\blue{The case $\beta=-1$.}} \blue{In this case,} we derive the limit law again using the method of moments. However, in contrast to the above proof, we \blue{do} not use generating functions, but instead use what is sometimes called the {\it moment-transfer approach}; see, e.g., \cite{Fu}.

First, note that we have (\ref{dis-rec}) with
\[
\pi_{n,j}=\frac{n}{2H_{n-1}}\cdot\frac{1}{j(n-j)},\qquad (1\leq j\leq n).
\]
Consequently, for the $m$-th moment:
\[
{\mathbb E}(D_{n,k}^m)=\frac{n}{H_{n-1}\binom{n}{k}}\sum_{j=1}^{n-1}\binom{j}{k}{\mathbb E}(D_{j,k}+1)^m\frac{1}{j(n-j)}
\]
or equivalently
\[
(n-1)_{(k-1)}{\mathbb E}(D_{n,k}^m)=\frac{1}{H_{n-1}}\sum_{j=1}^{n-1}\frac{(j-1)_{(k-1)}{\mathbb E}(D_{j,k}+1)^m}{n-j},
\]
where $n_{(k)}$ denotes the falling factorial. Using the binomial theorem gives:
\[
(n-1)_{(k-1)}{\mathbb E}(D_{n,k}^m)=\frac{1}{H_{n-1}}\sum_{j=1}^{n-1}\frac{1}{n-j}\sum_{\ell=0}^{m}\binom{m}{\ell}(j-1)_{(k-1)}{\mathbb E}(D_{j,k}^{\ell})
\]
or by setting $A_n^{[m]}:=(n-1)_{(k-1)}{\mathbb E}(D_{n,k}^{m})$ (from now on we suppress  the dependence on $k$):
\begin{equation}\label{rec-Anm}
A_n^{[m]}=\frac{1}{H_{n-1}}\sum_{j=1}^{n-1}\frac{A_j^{[m]}}{n-j}+\frac{1}{H_{n-1}}\sum_{\ell=0}^{m-1}\binom{m}{\ell}\sum_{j=1}^{n-1}\frac{A_{j}^{[\ell]}}{n-j}.
\end{equation}

This recurrence has the general form:
\begin{equation}\label{under-rec}
a_{n}=\frac{1}{H_{n-1}}\sum_{j=1}^{n-1}\frac{a_j}{n-j}+b_n
\end{equation}
for which the following asymptotic transfer result holds.

\bigskip

\begin{lem}\label{asymp-transfer} Let $t\in {\mathbb N}$ and $s\in{\mathbb Z}$.
\begin{itemize} 
\item[(i)] If $b_n={\mathcal O}(n^t\log^s n)$, then $a_n={\mathcal O}(n^{t}\log^{s+1}n)$.
\item[(ii)] If $b_n=cn^{t}\log^s n$, then $a_n=cn^{t}\log^{s+1}n/H_t+{\mathcal O}(n^{t}\log^{s}n)$.
\end{itemize}
\end{lem}
\begin{proof} This follows from the method introduced in \cite{ald24}. More precisely, in Section 2.11 of \cite{ald24}, Part (ii) was proved for $s=0$ and the same proof also contains Part (i) for $s=-1$. Generalizing the method, both claims of the lemma can be established in an analogous manner.\end{proof}

Using this result, we can now prove the following proposition.

\bigskip 

\begin{pro}
\label{Anm}
For $m\geq 1$, as $n\rightarrow\infty$,
\[
A_n^{[m]}=\frac{m!}{H_{k-1}^m}n^{k-1}\log^m n+{\mathcal O}(n^{k-1}\log^{m-1} n).
\]
\end{pro}

{\bf Remark:}
Proposition~2.14 in \cite{ald24} is the special case with $m=1$. Nevertheless, we below give a self-\blue{contained} proof also for this case.

\begin{proof} The proof proceeds from (\ref{rec-Anm}) by induction on $m$. First, for $m=1$, we have
\[
A_n^{[1]}=\frac{1}{H_{n-1}}\sum_{j=1}^{n-1}\frac{A_j^{[1]}}{n-j}+\frac{1}{H_{n-1}}\sum_{j=1}^{n-1}\frac{(j-1)_{(k-1)}}{n-j}.
\]
Note that 
\begin{align*}
\frac{1}{H_{n-1}}\sum_{j=1}^{n-1}\frac{(j-1)_{(k-1)}}{n-j}&=(n-1)_{(k-1)}\left(1-\frac{H_{k-1}}{H_{n-1}}\right)\\
&=n^{k-1}+{\mathcal O}(n^{k-1}\log^{-1}n).
\end{align*}
Thus, the claim follows from Lemma~\ref{asymp-transfer}.

Next, we assume that the induction claim is true for $m'$ with $m'<m$. We want to show it for $m$. We consider the term $\ell=m-1$ in the second term on the right-hand side of (\ref{rec-Anm}), i.e.,
\[
\frac{m}{H_{n-1}}\sum_{j=1}^{n-1}\frac{A_{j,k}^{[m-1]}}{n-j}.
\]
We now plug the induction assumption into this expression, where we first just use the main term. This gives
\begin{align*}
\frac{m!}{H_{k-1}^{m-1}H_{n-1}}&\sum_{j=1}^{n-1}\frac{j^{k-1}\log^{m-1}j}{n-j}\\
&=\frac{m!}{H_{k-1}^{m-1}H_{n-1}}\sum_{j=1}^{n-1}\frac{j^{k-1}
(\log(j/n)+\log n)^{m-1}}{n-j}\\
&=\frac{m!}{H_{k-1}^{m-1}H_{n-1}}\sum_{\ell=0}^{m-1}\binom{m-1}{\ell}(\log n)^{\ell}\sum_{j=1}^{n-1}\frac{j^{k-1}\log^{m-1-\ell}(j/n)}{n-j}.
\end{align*}
Note that for $t\in{\mathbb N}$:
\[
\sum_{j=1}^{n-1}\frac{j^{k-1}\log^t(j/n)}{n-j}\sim\left(\int_{0}^{1}\frac{x^{k-1}\log^{t}x}{1-x}{\rm d}x\right)n^{k-1}
\]
and
\[
\sum_{j=1}^{n-1}\frac{j^{k-1}}{n-j}=n^{k-1}\left(H_{n-1}+\sum_{j=1}^{n-1}\frac{(j/n)^{k-1}-1}{n-k}\right)=n^{k-1}H_{n-1}+{\mathcal O}(n^{k-1}).
\]
Overall,
\[
\frac{m!}{H_{k-1}^{m-1}H_{n-1}}\sum_{j=1}^{n-1}\frac{j^{k-1}\log^{m-1}j}{n-j}=\frac{m!}{H_{k-1}^{m-1}}n^{k-1}\log^{m-1}n+{\mathcal O}(n^{k-1}\log^{m-2} n).
\]
Likewise,
\[
\frac{m}{H_{n-1}}\sum_{j=1}^{n-1}\frac{{\mathcal O}(j^{k-1}\log^{m-2}j)}{n-j}={\mathcal O}(n^{k-1}\log^{m-2}n)
\]
and thus,
\[
\frac{m}{H_{n-1}}\sum_{j=1}^{n-1}\frac{A_{j,k}^{[m-1]}}{n-j}=\frac{m!}{H_{k-1}^{m-1}}n^{k-1}\log^{m-1}n+{\mathcal O}(n^{k-1}\log^{m-2} n).
\]
By the same argument:
\[
\frac{1}{H_{n-1}}\sum_{\ell=0}^{m-2}\binom{m}{\ell}\sum_{j=1}^{n-1}\frac{A_{j}^{[\ell]}}{n-j}={\mathcal O}(n^{k-1}\log^{m-2} n)
\]
and thus, the second term on the right-hand side of (\ref{rec-Anm}) becomes
\[
\frac{1}{H_{n-1}}\sum_{\ell=0}^{m-1}\binom{m}{\ell}\sum_{j=1}^{n-1}\frac{A_{j}^{[\ell]}}{n-j}=\frac{m!}{H_{k-1}^{m-1}}n^{k-1}\log^{m-1}n+{\mathcal O}(n^{k-1}\log^{m-2} n).
\]
Finally, applying (\ref{asymp-transfer}) gives the inductive claim.
\end{proof}

Proposition~\ref{Anm} now implies the following theorem, which proves the result claimed in Theorem~\ref{thm2}, Part (ii).

\bigskip

\begin{theorem}
As $n\rightarrow\infty$,
\[
\frac{H_{k-1}D_{n,k}}{\log n}\stackrel{d}{\longrightarrow}{\rm Exp}(1),
\]
where ${\rm Exp}(1)$ is the standard exponential distribution.
\end{theorem}
\begin{proof}
Recall that $A_n^{[m]}:=(n-1)_{(k-1)}{\mathbb E}(D_{n,k}^{m})$. Thus, from Proposition~\ref{Anm}, we have:
\[
{\mathbb E}(D_{n,k}^{m})\sim\frac{m!}{H_{k-1}^{m}}\log^m n
\]
or equivalently (with $X_{n,k}:=H_{k-1}D_{n,k}/\log n$):
\[
{\mathbb E}(X_{n,k}^{m})\sim m!.
\]
Since $m!$ are the moments of ${\rm Exp}(1)$ and this is the unique distribution with this moment sequence, \blue{the result follows by Lemma~\ref{bil}.}
\end{proof}

\subsection{Asymptotic geometric distribution for $\beta>-1$}\label{ll-geo}

We now show that for $\beta>-1$, the distribution of $D_{n,k}$ converges to a geometric distribution with success probability $1-q(\beta,k)$ as $n$ grows. The proof relies on the following lemma.

\bigskip

\begin{lem}
\label{lembal}
Let $E_n$ be the event that in a $\beta$-splitting tree with $n$ leaves, each of the two subtrees incident with the root has at least $\sqrt{n}$ leaves. Then, for $\beta> -1$, we have:
${\mathbb P}(E_n) \rightarrow 1$ as $n \rightarrow \infty.$
\end{lem}
\begin{proof} 
Let $I_n$ denote the size of the right subtree of the $\beta$-splitting tree; see the beginning of Section~\ref{proof-thm2}. It then suffices to show that ${\mathbb P}(I_n\leq\sqrt{n})\rightarrow 0$ as $n\rightarrow\infty$.

In order to show this, first, from (\ref{norm-constant}), we have for $\beta>-1$:
\[
c_n(\beta)\sim\frac{\Gamma(\beta+1)^2}{\Gamma(2\beta+2)}n^{2\beta+1}.
\]
Thus,
\begin{align*}
{\mathbb P}(I_n\leq \sqrt{n})= \sum_{1\leq j\leq \sqrt{n}} \pi_{n,j}&={\mathcal O}\left(n^{-2\beta-1}\sum_{j\leq\sqrt{n}}j^{\beta}(n-j)^{\beta} \right)\\
&={\mathcal O}\left(\int_{0}^{1/\sqrt{n}}x^{\beta}(1-x)^{\beta}{\rm d}x\right)
\end{align*}
which tends to $0$ as $n\rightarrow\infty$. This proves our claim.
\end{proof}

\begin{pro}
Suppose $\beta >-1$, and $k\geq 2$ is fixed. Then, as
 $n\rightarrow\infty$, 
\[
D_{n,k} \stackrel{d}{\longrightarrow} G_k,
\]
where $G_k$ has a geometric distribution, with 
${\mathbb P}(G_k = r) = (1-q(\beta, k))q(\beta, k)^{r}$ for $r \geq 0$.
\end{pro}

\begin{proof}
It suffices to show that for each $r \geq 1$, 
$$\lim_{n \rightarrow \infty}{\mathbb P}(D_{n,k} \geq r) = q(\beta,k)^r. $$
We use induction on $r$ starting with this base case $r=1$,
which holds by  (\ref{eqx}). For the induction step we use the following identity (which holds by definition):
\begin{equation}
\label{eqrecurs}
 {\mathbb P}(D_{n,k} \geq r+1) = {\mathbb P}(D_{n,k} \geq r+1|D_{n,k}\geq r)\cdot {\mathbb P}(D_{n,k} \geq r).   
\end{equation}
 To determine the first factor on the right of (\ref{eqrecurs}), we apply Lemma~\ref{lembal}, which implies that for any fixed value of $r\geq 1$, a $\beta$-splitting tree with $n$ leaves has the following two properties with a probability that converges to $1$ as $n$ grows:
(i) the tree has $2^r$ subtrees at depth $r$ from its root, and (ii)  each of these subtrees has at least $n^{2^{-r}}$ leaves.
Now, the event  $D_{n,k}\geq r$ is equivalent to the event that all of the $k$ randomly sampled leaves appear as leaves of just one of the $2^r$ trees that lie at depth $r$ from the root. Also,  each of these $2^r$ subtrees are described by the $\beta$-splitting model (with the same parameter value for $\beta$ as the original tree). Moreover, conditional on the event that $D_{n,k}\geq r$, the additional event that $D_{n,k}\geq r+1$ holds precisely if the $k$ leaves that lie within one subtree at depth $r$ from the root have a MRCA that has depth at least 1 to the root of that subtree.  Thus, by Lemma~\ref{lembal} and (\ref{eqx}),   $$\lim_{n \rightarrow \infty}{\mathbb P}(D_{n,k} \geq r+1|D_{n,k}\geq r) = \lim_{n \rightarrow \infty}{\mathbb P}(D_{n,k} \geq 1) = q(\beta,k),$$
which, combined with (\ref{eqrecurs}), establishes the induction step for $r$.
\end{proof}

\section{Concluding comments}

We end by pointing out two open problems.

First, a general observation concerning our results is that increasing $\beta$ leads to a higher probability that $D_{n,k}$ is small. Thus, it may be of interest to formally establish whether or not, for each value of $n, k$ and $r$, $\PP(D_{n,k} \leq r)$ is a monotone increasing function of $\beta$.

A further conjecture, suggested by using the same ansatz as in \cite{ald96}, is that for all $-2<\beta<-1$,
\[
\frac{D_{n,k}}{n^{-\beta-1}}\stackrel{d}{\longrightarrow} D_k,
\]
where $D_k$ is characterized by the moment sequence $\{c_m\}_{m\geq 1}$,
with
\[
c_m=m!\prod_{j=1}^{m}e(\beta,k,j).
\]
and
\[
e(\beta,k,m):=\frac{\Gamma((-\beta-1)m+\beta+k+1)}{\Gamma((-\beta-1)m+2\beta+k+2)}-\frac{\Gamma(\beta+2)}{\Gamma(2\beta+3)}.
\]
This holds for $\beta=-3/2$ (where $1/\Gamma(2\beta+3)=0$) as $c_m$ simplifies to the right-hand side of Equation (\ref{mom-Dk}). In order to prove the result more generally, tools similar to those in \cite{ald24} would have to be developed for the range $-2<\beta<-1$ which requires significant additional work. \blue{On the other hand, one could also try probabilistic tools, however, as the current paper relies mainly on combinatorial methods, such tools are outside of the scope of this work.}

\section{Acknowledgments} We thank Markus Kuba for pointing out that $D_k$ in Theorem~\ref{thm2} follows a Mittag-Leffler distribution; see the remark after Theorem~\ref{limit-law-subcrit}. \blue{We also thank Amaury Lambert for some helpful comments and suggestions.} This research was carried out during a sabbatical stay of MF at the Biomathematics Research Center, University of Canterbury, Christchurch. He thanks the department and Mike Steel for hospitality, and the National Science and Technology Council, Taiwan (research grants NSTC-113-2918-I-004-001 and NSTC-113-2115-M-004-004-MY3) for financial support. MS thanks the NZ Marsden Fund for research support (research grant 23-UOC-003).

\section{Data availability}
Data sharing is not applicable to this article as no datasets were generated or analysed during the current study.

\section*{Appendix: Random sampling of species}

In this appendix, we consider the following variation of the depth discussed throughout the main body of the paper: instead of fixing $k$ and picking a set of $k$ leaves uniformly at random, each leaf is now independently and at random included in a set with success probability $p$ (with the selection being repeated if the set is empty). 

Denote by $D_n$ the distance between the most recent common ancestor of this random  subset of leaves and  the root of the tree. We are going to derive similar results for ${\mathbb P}(D_n=0)$ to the results for ${\mathbb P}(D_{n,k})$ presented in Theorem~\ref{thm1}. 

Our starting point is the following lemma.

\bigskip

\begin{lem}\label{main-lemma}
We have,
\begin{equation}\label{Dn-0}
{\mathbb P}(D_n=0)=\frac{1-2{\mathbb E}(q^{I_n})+q^n}{1-q^n},
\end{equation}
where $I_n$ is a random variable with distribution $\pi_{n,j}$; see (\ref{split-probab}).
\end{lem}

\begin{proof}
Conditioning the probability we want to compute on $I_n=j$ gives
\[
{\mathbb P}(D_n=0\vert I_n=j)=\frac{(1-q^j)(1-q^{n-j})}{1-q^n}=\frac{1-q^j-q^{n-j}+q^n}{1-q^n}
\]
because $I_n$ is the number of points in $J$ (see the description of the $\beta$-splitting model in Section~\ref{def-models}) and $D_n=0$ if and only if not all sampled leaves are either chosen from $J$ or $n-J$. In addition, the denominator is because we require that the random set of sampled leaves is non-empty. Multiplying the above expression by $\pi_{n,j}$ and summing over $j$ gives the claimed result by the symmetry of $\pi_{n,j}$.
\end{proof}

We assume from now on that $pn=\lambda$ for a fixed $\lambda$, i.e., $p=\lambda/n$. Our main result in this appendix is as follows.

\bigskip

\begin{theorem}
\begin{itemize}\item[(i)] For $\beta>-1$, as $n\rightarrow\infty$,
\[
{\mathbb P}(D_n=0)\sim\frac{1-2\Gamma(2\beta+2)\int_{0}^{1}e^{-\lambda x}x^{\beta}(1-x)^{\beta}{\rm d}x/\Gamma(\beta+1)^2+e^{-\lambda}}{1-e^{-\lambda}}.
\]
\item[(ii)] For $\beta=-1$, as $n\rightarrow\infty$,
\[
{\mathbb P}(D_n=0)\sim\frac{E_1(\lambda)+\log\lambda+\gamma-e^{-\lambda}({\rm Ei}(\lambda)-\log\lambda-\gamma)}{(1-e^{-\lambda})H_{n-1}},
\]
where $E_1(z)$ and ${\rm Ei}(z)$ are the exponential integrals; see \cite{AbSt}.
\item[(iii)] For $-2<\beta<-1$, as $n\rightarrow\infty$,
\begin{align*}
{\mathbb P}(D_n=0)\sim\frac{n^{\beta+1}}{\Gamma(\beta+1)(1-e^{-\lambda})}\bigg(&\int_{0}^{1/2}(e^{-\lambda x}-1)x^{\beta}(1-x)^{\beta}{\rm d}x\\
&\qquad+\int_{1/2}^{1}(e^{-\lambda x}-e^{-\lambda})x^{\beta}(1-x)^{\beta}{\rm d}x\bigg).
\end{align*}
\end{itemize}
\end{theorem}

{\bf Remark:}
The asymptotic approximations above remain valid in the range $\lambda=o(\sqrt{n})$.

\begin{proof}
Due to Lemma~\ref{main-lemma}, the main task is to compute the expected value in (\ref{Dn-0}); the result then follows with
\[
q^n=\left(1-\frac{\lambda}{n}\right)^n\sim e^{-\lambda}
\]
which implies
\begin{equation}\label{Dn-00}
{\mathbb P}(D_n=0)\sim\frac{1-2{\mathbb E}(q^{I_n})+e^{-\lambda}}{1-e^{-\lambda}}.
\end{equation}

We consider the cases $\beta=-1$ and $\beta\ne -1$ separately, starting with the former, for which we have
\[
{\mathbb P}(I_n=j)=\frac{n}{2H_{n-1}}\cdot\frac{1}{j(n-j)}=\frac{1}{2H_{n-1}}\left(\frac{1}{j}+\frac{1}{n-j}\right).
\]
Consequently,
\[
{\mathbb E}(q^{I_n})=\frac{1}{2H_{n-1}}\left(\sum_{j=1}^{n-1}\frac{q^{j}}{j}+\sum_{j=1}^{n-1}\frac{q^{j}}{n-j}\right).
\]
Now,
\[
q^{j}=\left(1-\frac{\lambda}{n}\right)^j=e^{-j\lambda/n}\left(1+{\mathcal O}\left(\frac{1}{n}\right)\right),
\]
where this holds uniformly for $1\leq j\leq n$. Thus,
\begin{align*}
{\mathbb E}(q^{I_n})&\sim\frac{1}{2H_{n-1}}\left(\sum_{j=1}^{n-1}\frac{e^{-j\lambda/n}}{j}+\sum_{j=1}^{n-1}\frac{e^{-j\lambda/n}}{n-j}\right)\\
&=\frac{1}{2H_{n-1}}\left(H_{n-1}+\sum_{j=1}^{n-1}\frac{e^{-j\lambda/n}-1}{j}+e^{-\lambda}H_{n-1}
+\sum_{j=1}^{n-1}\frac{e^{-j\lambda/n}-e^{-\lambda}}{n-j}\right)\\
&\sim\frac{1+e^{-\lambda}}{2}+\frac{1}{2H_{n-1}}\left(\int_0^1\frac{e^{-\lambda x}-1}{x}{\rm dx}+\int_{0}^{1}\frac{e^{-\lambda x}-e^{-\lambda}}{1-x}{\rm d}x\right).
\end{align*}
Note that
\[
\int_0^1\frac{e^{-\lambda x}-1}{x}{\rm dx}=\int_{0}^{\lambda}\frac{e^{-t}-1}{t}{\rm d}t=-E_1(\lambda)-\log\lambda-\gamma,
\]
where $E_1(z)$ denotes the exponential integral and $\gamma$ is Euler's constant. Also,
\[
\int_{0}^{1}\frac{e^{-\lambda x}-e^{-\lambda}}{1-x}{\rm d}x=e^{-\lambda}\int_0^{\lambda}\frac{e^{t}-1}{t}{\rm d}t=e^{-\lambda}({\rm Ei}(\lambda)-\log\lambda-\gamma),
\]
where ${\rm Ei}(z)$ is another of the exponential integrals. Thus,
\[
{\mathbb E}(q^{I_n})\sim\frac{1+e^{-\lambda}}{2}-\frac{E_1(\lambda)+\log\lambda+\gamma-e^{-\lambda}({\rm Ei}(\lambda)-\log\lambda-\gamma)}{2H_{n-1}}
\]
and plugging this into (\ref{Dn-00}) gives the claim for $\beta=-1$.

Next, we consider $\beta\ne -1$, for which we have
\begin{align*}
{\mathbb P}(I_n=j)&=\frac{1}{c_n(\beta)}\frac{\Gamma(j+\beta+1)\Gamma(n-j+\beta+1)}{j!(n-j)!}\\
&=\frac{\Gamma(\beta+1)^2}{c_n(\beta)}\binom{j+\beta}{j}\binom{n-j+\beta}{\beta}\\
&=\frac{1}{\binom{n+2\beta+1}{n}-2\binom{n+\beta}{n}}\binom{j+\beta}{j}\binom{n-j+\beta}{n-j};
\end{align*}
see (\ref{norm-constant}) for the last step. Since the denominator has different asymptotics if $\beta>-1$ and $-2<\beta<-1$ (see (\ref{diff-asymp})), we consider now these two cases separately.

First, for $\beta>-1$, we have
\begin{align*}
{\mathbb E}(q^{I_n})&\sim\frac{\Gamma(2\beta+2)}{\Gamma(\beta+1)^2n^{2\beta+1}}\sum_{j=1}^{n-1}e^{-j\lambda/n}j^{\beta}(n-j)^{\beta}\\
&\sim\frac{\Gamma(2\beta+2)}{\Gamma(\beta+1)^2}\int_{0}^{1}e^{-\lambda x}x^{\beta}(1-x)^{\beta}{\rm d}x.
\end{align*}
Plugging this into (\ref{Dn-00}) gives the claimed result in this case.

On the other hand, for $-2<\beta<-1$, we have
\[
{\mathbb E}(q^{I_n})\sim\frac{\Gamma(\beta+1)}{-2n^{\beta}}\sum_{j=1}^{n-1}e^{-j\lambda/n}\binom{j+\beta}{j}\binom{n-j+\beta}{n-j}.
\]
We split the sum into two parts according to whether $1\leq j\leq (n-1)/2$ or $(n-1)/2<j\leq n-1$. For the first part, we have
\begin{align*}
\sum_{j=1}^{(n-1)/2}&e^{-j\lambda/n}\binom{j+\beta}{j}\binom{n-j+\beta}{n-j}\\[4pt]
&=\sum_{j=1}^{(n-1)/2}\binom{j+\beta}{j}\binom{n-j+\beta}{n-j}+
\sum_{j=1}^{n-1}(e^{-j\lambda/n}-1)\binom{j+\beta}{j}\binom{n-j+\beta}{n-j}\\[4pt]
&\sim\frac{1}{2}\sum_{j=1}^{(n-1)/2}\binom{j+\beta}{j}\binom{n-j+\beta}{n-j}+\frac{n^{2\beta+1}}{\Gamma(\beta+1)^2}\int_{0}^{1/2}(e^{-\lambda x}-1)x^{\beta}(1-x)^{\beta}{\rm d}x\\[4pt]
&\sim-\frac{n^{\beta}}{\Gamma(\beta+1)}+\frac{n^{2\beta+1}}{\Gamma(\beta+1)^2}\int_{0}^{1/2}(e^{-\lambda x}-1)x^{\beta}(1-x)^{\beta}{\rm d}x.
\end{align*}
Likewise,
\begin{align*}
\sum_{(n-1)/2<j\leq n-1}e^{-j\lambda/n}&\binom{j+\beta}{j}\binom{n-j+\beta}{n-j}\\
&\sim-\frac{e^{-\lambda}n^{\beta}}{\Gamma(\beta+1)}+\frac{n^{2\beta+1}}{\Gamma(\beta+1)^2}
\int_{1/2}^{1}(e^{-\lambda x}-e^{-\lambda})x^{\beta}(1-x)^{\beta}{\rm d}x.
\end{align*}
Thus,
\begin{align*}
{\mathbb E}(q^{I_n})\sim\frac{1+e^{-\lambda}}{2}&-\frac{n^{\beta+1}}{2\Gamma(\beta+1)}\bigg(\int_{0}^{1/2}(e^{-\lambda x}-1)x^{\beta}(1-x)^{\beta}{\rm d}x\\
&+\int_{1/2}^{1}(e^{-\lambda x}-e^{-\lambda})x^{\beta}(1-x)^{\beta}{\rm d}x\bigg)
\end{align*}
and plugging this into (\ref{Dn-00}) gives the final claim of the theorem.
\end{proof}

{\bf Remark:}
If we let $\lambda\rightarrow\infty$ in the asymptotic expansions of ${\mathbb P}(D_n=0)$, we recover the results from Theorem~\ref{thm1} with $k\rightarrow\infty$. We explain this in this remark for $\beta=-1$, $\beta>-1$, and $-2<\beta<-1$ separately.

First, for $\beta=-1$, as $\lambda\rightarrow\infty$,
\[
E_1(\lambda)+\log\lambda+\gamma-e^{-\lambda}({\rm Ei}(\lambda)-\log\lambda-\gamma)\sim\log\lambda
\]
and thus,
\[
\frac{E_1(\lambda)+\log\lambda+\gamma-e^{-\lambda}({\rm Ei}(\lambda)-\log\lambda-\gamma)}{(1-e^{-\lambda})H_{n-1}}\sim\frac{\log\lambda}{H_{n-1}}
\]
which matches the asymptotics of the expression on the right-hand of (\ref{beta=-1}).

Next, for $\beta>-1$, as $\lambda\rightarrow\infty$,
\[
\int_{0}^{1}e^{-\lambda x}x^{\beta}(1-x)^{\beta}{\rm d}x=\frac{1}{\lambda^{\beta+1}}\int_0^{\lambda}e^{-t}t^{\beta}(1-t/\lambda)^{\beta}{\rm d}\sim\frac{\beta+1}{\Gamma(\lambda^{\beta+1})}
\]
and thus,
\[
\frac{1-2\Gamma(2\beta+2)\int_{0}^{1}e^{-\lambda x}x^{\beta}(1-x)^{\beta}{\rm d}x/\Gamma(\beta+1)^2+e^{-\lambda}}{1-e^{-\lambda}}\sim 1-\frac{2\Gamma(2\beta+2)}{\lambda^{\beta+1}\Gamma(\beta+1)}.
\]
This again matches with the case of fixed $k$ (see Theorem~\ref{thm1}, Part (ii)) since, as $k\rightarrow\infty$,
\begin{align*}
q(\beta,k)=\frac{(\beta+2)\cdots(\beta+k)}{(2\beta+3)\cdots (2\beta+k+1)}&=\frac{2\Gamma(2\beta+2)\Gamma(\beta+k+1)}{\Gamma(\beta+1)\Gamma(2\beta+k+2)}\\
&\sim\frac{2\Gamma(2\beta+2)}{k^{\beta+1}\Gamma(\beta+1)}.
\end{align*}

Finally, for $-2<\beta<-1$, the two integrals in the asymptotics of ${\mathbb P}(D_n=0)$ become
\[
\int_{0}^{1/2}(e^{-\lambda x}-1)x^{\beta}(1-x)^{\beta}{\rm d}x=\frac{1}{\lambda^{\beta+1}}\int_{0}^{\lambda/2}(e^{-t}-1)t^{\beta}(1-t/\lambda)^{\beta}{\rm d}t\sim\frac{\Gamma(\beta+1)}{\lambda^{\beta+1}}
\]
and
\[
\int_{1/2}^{1}(e^{-\lambda x}-e^{-\lambda})x^{\beta}(1-x)^{\beta}{\rm d}x=e^{-\lambda}\int_{0}^{1/2}(e^{\lambda x}-1)x^{\beta}(1-x)^{\beta}{\rm d}x={\mathcal O}(e^{-\lambda/2}).
\]
Thus, as $\lambda\rightarrow\infty$,
\begin{align*}
\frac{n^{\beta+1}}{\Gamma(\beta+1)(1-e^{-\lambda})}&\left(\int_{0}^{1/2}(e^{-\lambda x}-1)x^{\beta}(1-x)^{\beta}{\rm d}x+\int_{1/2}^{1}(e^{-\lambda x}-e^{-\lambda})x^{\beta}(1-x)^{\beta}{\rm d}x\right)\\
&\sim\left(\frac{\lambda}{n}\right)^{-\beta-1}
\end{align*}
({\em cf.} Equation \ref{beta<-1}).

\end{document}